\documentclass[prb, aps, onecolumn, showpacs]{revtex4}
\usepackage{amssymb}
\usepackage{amsmath}
\usepackage[dvips]{graphicx}
\usepackage[english]{babel}

\begin{document}

\title{Electron-hole symmetry and solutions of Richardson pairing model}
\author{W. V. Pogosov$^{1,2}$, N. S. Lin$^{1}$, and V. R. Misko$^{1}$}
\affiliation{$^{1}$Departement Fysica, Universiteit Antwerpen,
Groenenborgerlaan 171, B-2020 Antwerpen, Belgium}
\affiliation{$^{2}$Institute for Theoretical and Applied
Electrodynamics, Russian Academy of Sciences, Izhorskaya 13, 125412,
Moscow, Russia}

\begin{abstract}
Richardson approach provides an exact solution of the pairing
Hamiltonian. This Hamiltonian is characterized by the electron-hole
pairing symmetry, which is however hidden in Richardson equations.
By analyzing this symmetry and using an additional conjecture,
fulfilled in solvable limits, we suggest a simple expression of the
ground state energy for an equally-spaced energy-level model, which
is applicable along the whole crossover from the superconducting
state to the pairing fluctuation regime. Solving Richardson
equations numerically, we demonstrate a good accuracy of our
expression.
\end{abstract}

\author{}
\maketitle
\date{\today }

\section{Introduction}

As shown by Richardson, the ``reduced'' Hamiltonian of the
Bardeen-Cooper-Schrieffer (BCS) theory of superconductivity is
exactly solvable\cite{Rich1}. The Richardson approach is based on
the canonical ensemble, i.e., the number of particles is fixed,
while the BCS theory corresponds to the grand canonical description,
which becomes accurate in the large-sample limit. In addition, the
BCS theory provides only a mean-field approximation, which
nevertheless turns out to be exact\cite{Rich3} in the large-sample
limit due to peculiarities of the ``reduced'' BCS potential.
However, the grand canonical BCS theory is definitely not applicable
for small-sized systems accommodating few pairs only. Significant
improvements, nevertheless, are possible if one incorporates
canonical approach into the BCS model (particle-number projected
BCS), but still a rather heavy numerics is needed to proceed with
computations\cite{Braun}. For the applicability of the canonical
ensemble to the theory of superconductivity see Ref.
\cite{Bogoliubov1}.

In contrast to the BCS theory, the Richardson method yields an exact
solution to the many-body problem involving BCS pairing Hamiltonian.
Within this approach, the energy of the system of $N$ correlated
pairs is expressed through the sum of $N$ energy-like quantities,
which satisfy the system of $N$ coupled nonlinear equations, called
Richardson equations. It is remarkable that they are also derivable
through the algebraic Bethe-ansatz approach\cite{Pogosyan}, so that
the Richardson equations can be considered as one of the examples of
Bethe-ansatz equations. However, it turns out that solving the
Richardson equations is a formidable task. Up to now, they have been
evaluated explicitly only in few special cases. In particular, no
analytical solution exists for the crossover between the
superconducting state and the pairing fluctuation regime, which is
relevant for small-sized systems even for zero temperature. In this
situation, no small parameter seems to exist, which could be used to
construct some expansion. For this case, the Richardson equations
are tackled numerically, when studying small systems described by
pairing Hamiltonian, among which are nanosized superconducting
grains~\cite{Dukel}, nuclei~\cite{Dukel1}, ultracold
atoms~\cite{Belzig}, bubbles in liquid helium~\cite{Gladilin1} (for
studies of correlation functions, see Ref.~\cite{Osterloh}).

The aim of the present paper is to apply \textit{symmetry arguments}
in order to provide analytical results for the crossover region. It
is actually well known that symmetry considerations can be very
helpful in situations, when brute-force methods are not so
efficient. We show that the BCS pairing Hamiltonian has an
electron-hole (e-h) symmetry and we then try to reveal the impact of
this symmetry for the ground state energy for the arbitrary number
of pairs and interaction constant. The case of the equally-spaced
model is considered, when the energy levels of noninteracting
particles are distributed equidistantly. The impact of e-h symmetry
can be most readily revealed in this particular case.

Although the e-h symmetry is encoded in the pairing Hamiltonian, it
does not show up explicitly in Richardson equations. From the
symmetry arguments, we suggest a simple explicit formula for the
ground state energy, which constitutes a main result of this paper.
Within our approach, we use a conjecture on the $N$-dependence of
the dominant contribution to the ground state energy (when all other
parameters are fixed). This hypothesis is justified by the fact that
a guessed simple dependence on $N$ emerges in very different
analytically-solvable limits. Under this assumption, we actually
reduce the problem to the resolution of a \textit{single} Richardson
equation, which is far simpler than the solution of the full set of
equations. The resulting expression for the ground state energy is
also in a perfect agreement with known results in these limits,
although we do \textit{not} solve the whole system of Richardson
equations. In order to address the accuracy of our formula in the
crossover regime, we solve numerically the full system of these
equations for $N$ pairs changing from $N=1$ to 50 and compare the
results. We found a very good accuracy for systems with small number
of pairs, which are of particular interest due to limitations of BCS
treatment in this case. For systems with larger number of pairs, the
maximum error, which corresponds to the regime of intermediate
strength of coupling, grows. Nevertheless, our expression remains
applicable in this case too. Our approach can be useful for other
types of Bethe-ansatz equations, especially for those, which
correspond to Gaudin-like models.

Though BCS Hamiltonian is rather simple, its applicability to small
diffusive superconducting grains has been proved, see, e.g., Refs.
\cite{Kurland,Aleiner,Ralph}. The crucial condition is that the
Thouless energy, which gives the inverse time to diffuse across the
grain~\cite{Vinokur}, must be much larger than the average energy
level spacing\cite{Kurland}. Also important is to have the Thouless
energy much larger than the superconducting gap (see Appendix B of
Ref. \cite{Aleiner}). It is less obvious what should be a relation
between the energy level spacing and the superconducting gap. In
particular, it was argued in Ref. \cite{Ralph} (see also Ref.
\cite{Aleiner}) that the BCS Hamiltonian probably can be considered
as a toy model only, when the gap becomes much smaller than the
level spacing.

\section{Model}

We consider a system of fermions of two sorts, for instance, with
spins up and down. Particles attract each other through the BCS
``reduced'' potential, coupling only fermions of different sorts and
with zero total momenta as
\begin{equation}
\mathcal{V}=-V\sum_{\mathbf{k},\mathbf{k}^{\prime
}}a_{\mathbf{k}^{\prime
}\uparrow }^{\dagger }a_{-\mathbf{k}^{\prime }\downarrow }^{\dagger }a_{-%
\mathbf{k}\downarrow }a_{\mathbf{k}\uparrow }.  \label{BCSpot}
\end{equation}%
The total Hamiltonian is $H=H_{0}+\mathcal{V}$, where
\begin{equation}
H_{0}=\sum_{\mathbf{k}}\varepsilon _{\mathbf{k}}\left(
a_{\mathbf{k}\uparrow
}^{\dagger }a_{\mathbf{k}\uparrow }+a_{\mathbf{k}\downarrow }^{\dagger }a_{%
\mathbf{k}\downarrow }\right).  \label{H0}
\end{equation}
The summation in the right-hand side (RHS) of Eq. (\ref{BCSpot})
runs only over the states with
kinetic energies $\varepsilon _{\mathbf{k}}$ and $\varepsilon _{\mathbf{k}%
^{\prime }}$ located in the energy band between $\varepsilon
_{F_{0}}$\ and \ $\varepsilon _{F_{0}}+\Omega $ (Debye window).
These energies are distributed equidistantly (equally-spaced model),
so that the difference between two nearest values of $\varepsilon
_{\mathbf{k}}$ is $1/ \rho$. The density of states $ \rho$ increases
with the system volume, while the interaction constant $V$
decreases, so that the dimensionless interaction constant $v=\rho V$
in superconductors is finite and, in the large-sample limit, it can
be treated as a material characteristics. In the BCS theory, the
energy interval between $\varepsilon _{F_{0}}$\ and \ $\varepsilon
_{F_{0}}+\Omega $ is assumed to be always half-filled, while
$\Omega/2 $ is the Debye frequency. Thus, the
total number of available states with up or down spins in the Debye window is $%
N_{\Omega }\equiv \rho \Omega $, while $N=N_{\Omega }/2$;
$\varepsilon _{\mathbf{k}}$ runs from $\varepsilon _{F_{0}}$ to
$\varepsilon _{F_{0}}+(N_{\Omega }-1)/ \rho $ taking $N_{\Omega }$
values in total.

While it is supposed that the half-filling configuration only is
physically meaningful, one can consider other fillings and at least
treat the problem from the purely mathematical perspective. Note
however that it was argued in Ref. \cite{Geyer} that such a model
might be relevant to some semiconductors (see also Ref.
\cite{Eagles}). Introduction of the extra degree of freedom, which
is a filling, is an important ingredient of our analysis. By
considering the energy of the system formally as a function of $N$
(with all other input parameters fixed), we are going to obtain a
valuable information on the half-filling situation. Numerical
results will be presented for $N=N_{\Omega }/2$ only. For the sake
of simplicity, we will focus on even values of $N_{\Omega }$.

In the present paper, we concentrate on the equally-spaced model,
which provides a simplest but physically meaningful distribution of
energy levels. Therefore, this model is the most attractive starting
point to study the impact of e-h symmetry on the solutions of
Richardson model. Superconducting correlations were also studied in
finite-size system with random spacings of levels, distributed in
accordance with the gaussian orthogonal ensemble, using
mean-field\cite{Ambeg} and exact Richardson approaches\cite{Dussel}.
Such a statistics is typical for small metallic grains\cite{Gorkov}.
It was shown~\cite{Dussel} that the crossover between the
superconducting state and the fluctuation-dominated regime is
smooth, similarly to the case of the equally-spaced model.

The Hamiltonian, defined in Eqs. (\ref{BCSpot}) and (\ref{H0}), is
exactly solvable\cite{Rich1}. The energy of $N$ pairs is given by
the sum of $N$ rapidities $R_{j}$ ($j=1$,..., $N$) as
$E_{N}=\sum_{j}R_{j}$. The Richardson equation for each rapidity $R_{j}$ reads%
\begin{equation}
1=\sum_{\mathbf{k}}\frac{V}{2\varepsilon
_{\mathbf{k}}-R_{j}}+\sum_{l,l\neq j}\frac{2V}{R_{j}-R_{l}},
\label{Richardson}
\end{equation}
where the summation in the first term is performed over $\varepsilon
_{\mathbf{k}}$ located in the Debye window. Note that the dependence
of $E_{N}$ on $N$ thus enters through the number of equations.

\section{The electron-hole symmetry}

Now we discuss the internal electron-hole symmetry contained in the
Hamiltonian, from which we are going to deduce an information on the
solutions of Richardson equations. Let us introduce creation
operators for holes as $b_{\mathbf{k}\uparrow
}^{\dagger}=a_{\mathbf{k}\uparrow }$ and $b_{\mathbf{k}\downarrow
}^{\dagger}=a_{\mathbf{k}\downarrow }$. By using commutation
relations for fermionic operators, it is easy to
rewrite the Hamiltonian in terms of holes as%
\begin{eqnarray}
H=-VN_{\Omega}+&2&\sum_{\mathbf{k}}\varepsilon _{\mathbf{k}}-\sum_{%
\mathbf{k}}\left( \varepsilon _{\mathbf{k}}-V\right) \left( b_{\mathbf{k}%
\uparrow }^{\dagger }b_{\mathbf{k}\uparrow }+b_{\mathbf{k}\downarrow
}^{\dagger }b_{\mathbf{k}\downarrow
}\right)
\nonumber \\
-&V&\sum_{\mathbf{k},\mathbf{k}%
^{\prime }}b_{\mathbf{k}^{\prime }\uparrow }^{\dagger }b_{-\mathbf{k}%
^{\prime }\downarrow }^{\dagger }b_{-\mathbf{k}\downarrow }b_{\mathbf{k}%
\uparrow }. \label{BCSholes}
\end{eqnarray}%
The first two terms of the RHS of Eq. (\ref{BCSholes}) are numbers.
They give the potential energy and the kinetic energy of the Debye
window completely filled by electron pairs, respectively. The fourth
term coincides precisely with the interaction
potential in terms of electrons, given by Eq. (%
\ref{BCSpot}). To analyze the third term, we introduce $\xi
_{\mathbf{k}}^{^{\prime }}$, defined as $\xi _{\mathbf{k}}^{^{\prime
}}=$ $\varepsilon _{F_{0}}+(N_{\Omega }-1)/ \rho -\varepsilon
_{\mathbf{k}}$, which takes values $0$, $1/\rho $, $2/\rho $, ...,
$(N_{\Omega }-1)/ \rho$, so that $\xi _{\mathbf{k}}^{^{\prime }}$
runs over all the states starting from the top of the Debye window
towards its bottom,
i.e., in the inverse order. Then, $%
-(\varepsilon _{\mathbf{k}}-V$) can be represented as $\xi _{\mathbf{k}%
}^{^{\prime }}+(V- \varepsilon _{F_{0}}-(N_{\Omega }-1)/ \rho)$. A
similar term in the
Hamiltonian for electrons, given by Eq. (\ref{H0}), contains a factor $\xi _{%
\mathbf{k}}+\varepsilon _{F_{0}}$, where $\xi _{ \mathbf{k}}$ takes
values $0$, $1/\rho $, $2/\rho $, ..., $(N_{\Omega }-1)/ \rho$, so
that it runs over all states starting from the bottom to the top of
the same energy band.

Thus, due to peculiarities of the interaction potential, there
exists a symmetry between electron and hole pairs in the
Hamiltonian. Moreover, due to the equally-spaced distribution of
energy levels in the Debye window, the ground state energy of $N$
electron pairs can be explicitly expressed through the energy of
$N_{\Omega }-N$ electron pairs, with $\varepsilon _{F_{0}}$ changed
into $(V- \varepsilon _{F_{0}}-(N_{\Omega }-1)/ \rho)$. We therefore
treat this energy $E_{N}(\varepsilon _{F_{0}})$ as a function of
$\varepsilon _{F_{0}}$, which is an arbitrary nonzero number, and a
discrete variable $N$, which runs over the set $1, 2,..., N_{\Omega
}$. Since the bare kinetic energy of a filled Debye window is given
by the sum of terms of the arithmetic progression,
$2N_{\Omega}\varepsilon _{F_{0}}+N_{\Omega }(N_{\Omega }-1)/ \rho$,
we arrive at the identity
\begin{eqnarray}
E_{N}(\varepsilon _{F_{0}})=E_{N_{\Omega }-N}\left( V- \varepsilon
_{F_{0}}-\frac{N_{\Omega }-1}{\rho}\right)
\nonumber \\
-V N_{\Omega } +2N_{\Omega }\varepsilon _{F_{0}} + \frac{N_{\Omega
}(N_{\Omega }-1)}{ \rho}. \label{switch}
\end{eqnarray}
Note that in the case of more complex types of energy-level
distributions in the conduction band, one has also to change the
distribution, when switching from electron pairs to hole pairs,
which corresponds to counting levels from the top instead of
counting them from the bottom of the Debye window. This makes the
situation more subtle compared to the equally-spaced model.
Nevertheless, the duality between the electrons and holes still
exists, since this feature is a direct consequence of BCS
interaction potential, so that Eq. (\ref{BCSholes}) stays valid.

Next, we split $E_{N}(\varepsilon _{F_{0}})$ into the additive
contribution $2N\varepsilon _{F_{0}}$, which simply corresponds to
the shift of all rapidities, and $E_{N}^{'}$, the latter being
independent of $\varepsilon _{F_{0}}$
\begin{equation}
E_{N}(\varepsilon _{F_{0}})=2N\varepsilon _{F_{0}}+E_{N}^{'}.
\label{split}
\end{equation}
By substituting Eq. (\ref{split}) to (\ref{switch}), we arrive at
the functional equation
\begin{equation}
E_{N}^{'}=E_{N_{\Omega }-N}^{'} +(N_{\Omega
}-2N)\left(V-\frac{N_{\Omega }-1}{\rho}\right), \label{switch-cond}
\end{equation}
which is still exact. This remarkable condition will enable us to
relate the condensation energy of $N$ pairs to the condensation
energy of $N_{\Omega }-N$ pairs. Note that it is automatically
fulfilled for the half-filling, $N=N_{\Omega }/2$.

We would like to stress that, although the e-h symmetry can be
rather easily extracted from the Hamiltonian, it is not obvious that
the solution of $N$ equations is related to that of $N_{\Omega} - N$
equations through such a simple and universal relation as
Eq.~(\ref{switch-cond}). For the moment, we do not know how this
relation can be derived directly from Richardson equations.

Since $E_{N}^{'}$ is a function of the discrete variable $N$, which
runs over $N_{\Omega }$ values, it can always be represented as a
polynomial of $N$ of power $N_{\Omega }$. Alternatively, instead of
expanding in elementary monomials $N^{n}$, one may use Pochhammer
symbols defined as
\begin{equation}
(N)_{n}=N(N-1)...(N-n+1),  \label{pochh}
\end{equation}%
while $(N)_{0}\equiv 1$; so that $(N)_{n}$ may be treated as a
polynomial of $N$ of power $n$. Then,
\begin{equation}
E_{N}^{'}=\sum_{n=1}^{N_{\Omega }}e_{n}(N)_{n}, \label{alfa-sums}
\end{equation}%
where {$e_{n}$} is a set of unknown numbers.

Actually, $E_{N}^{'}$ can be also split into the condensation energy
$E_{N}^{(cond)}$ and the contribution coming from the bare kinetic
energy. The latter is given universally by $N(N-1)/\rho$, which can
be obviously described by the second term in the RHS of Eq.
(\ref{alfa-sums}).

Up to now, all the results were exact. At this step, we make a
conjecture that a dominant contribution to $E_{N}^{'}$ is due to the
first two terms in the sum of the RHS of Eq. (\ref{alfa-sums}) that
is
\begin{equation}
E_{N}^{'}\simeq e_{1}N+e_{2}N(N-1). \label{alfa-app}
\end{equation}%
This assumption is fully reasonable, since such a form of
$E_{N}^{'}$ does emerge in three important limits, which are
solvable analytically. Let's discuss the condensation energy in
these three limits, since the contribution from the kinetic energy,
$N(N-1)/\rho$, is always in agreement with Eq. (\ref{alfa-app}), as
discussed above. The first limit is a regime of the very weak
coupling realized for finite $N_{\Omega }$ ($v \ll 1/\ln (N_{\Omega
})$), for which all the rapidities are located in real axis and
approach the energy levels of noninteracting electrons. In this
case, $E_{N}^{(cond)}=-VN$, so that Eq. (\ref{alfa-app}) is
satisfied. Another limit is the strong-coupling regime ($v \gg 1$),
when all the rapidities are located far away from the line of
one-electron levels in the complex plane. In this case,
$E_{N}^{(cond)}$ contains\cite{Rich1} terms proportional to $N$ and
$N(N-1)$. At last, there is a limit of infinite $N$ at finite
nonzero $v$; here also $E_{N}^{(cond)}$ has a similar
form\cite{Rich3,We}. These three limits are quite different from
each other\cite{Dussel,Schechter,Altshuler}. It is actually the
reason why we conjecture that the structure of the solution, given
by Eq. (\ref{alfa-app}), must remain robust in the intermediate
region, which is characterized by finite $N_{\Omega}$ and arbitrary
$v$.

Now we consider $e_{1}$ and $e_{2}$ as unknown numbers and
substitute Eq. (\ref{alfa-app}) into Eq. (\ref{switch-cond}). We
then equate coefficients of $(N)_{0}$, $(N)_{1}$ and $(N)_{2}$ in
both sides of this equation and obtain a system of three linear
equations for $e_{1}$ and $e_{2}$. These three equations turn out to
be dependent, so they yield only a single condition as
\begin{equation}
e_{2}=-\frac{e_{1}}{N_{\Omega }-1}+\left(\frac{N_{\Omega
}-1}{\rho}-V\right)\frac{1}{N_{\Omega }-1}. \label{alfa2}
\end{equation}%
At this stage, we are left with only one unknown number, $e_{1}$. It
can be easily determined by considering a one-pair problem (as a
``boundary condition'' in the space of discrete $N$). In this case,
$E_{1}^{'}=e_{1}$, while $e_{1}$ is given by the solution of the
single Richardson equation
\begin{equation}
V=\sum_{n=0}^{N_{\Omega }-1}\frac{1}{2n/\rho - e_{1}},
\label{Cooper}
\end{equation}
under the condition $e_{1}<0$, which ensures that we select the
lowest-energy solution; then, $-e_{1}$ is a binding energy of a
single pair. In the general case, $e_{1}$ has to be determined
numerically, while exact analytical results are available in certain
limits, as discussed below. Note that Eq. (\ref{Cooper}) can be
rewritten in terms of $\Gamma$-functions.

The expression of $E_{N}^{'}$ is obtained by substituting Eq.
(\ref{alfa2}) to Eq. (\ref{alfa-app}) as
\begin{equation}
E_{N}^{'}=\frac{N(N-1)}{\rho}+N e_{1}\left(1-\frac{N-1}{N_{\Omega
}-1}\right)-VN \frac{N-1}{N_{\Omega }-1}, \label{final}
\end{equation}
where the first term comes from the bare kinetic energy, while two
others give the condensation energy. It is remarkable that our
method reproduces the first contribution automatically in the exact
form. The condensation energy per pair then reads
\begin{equation}
E_{N}^{(cond)}/N=e_{1}\left(1-\frac{N-1}{N_{\Omega }-1}\right)-V
\frac{N-1}{N_{\Omega }-1}. \label{final-cond}
\end{equation}
Eq. (\ref{final-cond}) supplemented by Eq. (\ref{Cooper}) is the
main result of this paper.

\section{Discussion}

Let us now consider the limits when Eq.~(\ref{Cooper}) can be solved
analytically, in order to see whether Eq.~(\ref{final-cond}) gives
reasonable results.

We start with the limit of the very weak coupling, in which the
solution of Eq. (\ref{Cooper}) approaches the lowest level so
closely that the mutual separation becomes much smaller than
$1/\rho$ and, moreover, contributions from other levels can be
neglected. Then, we obtain a simple solution $e_{1}=-V$. By
estimating dropped contributions due to other levels, we obtain a
criterion of applicability of this result as $v\ll 1/\ln(N_{\Omega
})$. In this case, $E_{N}^{(cond)}/N$ reduces to $-V$, as it must
be, due to the cancellation in the RHS of Eq. (\ref{final-cond}).
Note that the condensation energy per pair in this limit is
independent of filling.

Next, we consider an opposite limit, when the separation between the
single rapidity and the lowest level is much larger than $1/\rho$
and the number of levels $N_{\Omega }$ is also large. This enables
us to replace the sum in the RHS of Eq. (\ref{Cooper}) by the
integral, which gives
\begin{equation}
e_{1}\simeq-2\Omega\frac{\exp(-2/v)}{1-\exp(-2/v)}. \label{binding}
\end{equation}
The condition of applicability of this result is thus twofold:
$N_{\Omega }\gg 1$ and $N_{\Omega }\exp(-2/v)/(1-\exp(-2/v))\gg 1$.
The infinite-sample limit, when $v$ is fixed and finite, thus always
satisfies these criteria. It is also easy to see that $ \mid e_{1}
\mid $ given by Eq.~(\ref{Cooper}) is much larger than $V$ in the
large-sample limit, $N_{\Omega }\longrightarrow \infty$, so that
$V\sim \mid e_{1} \mid/N_{\Omega }$. This means that in this limit
$V$ can be neglected in Eq.~(\ref{final-cond}). Then,
\begin{equation}
E_{N}^{(cond)}/N\simeq -2\left(\Omega-\frac{N}{\rho}\right)
\frac{\exp(-2/v)}{1-\exp(-2/v)}, \label{final-BCS}
\end{equation}
which, for the half-filling, coincides with the BCS expression for
the condensation energy per $N$ (and with Richardson large-$N$
result for the same filling\cite{Rich3}). For arbitrary filling, it
also coincides with the results of both the mean-field treatment and
of the Richardson approach\cite{We}.

We would like to stress that the obtained results are highly
nontrivial. Indeed, the condensation energy given by Eq.
(\ref{final-cond}) consists of two terms. In the limit of a very
weak coupling, both of them are of the same order, while their
combination gives the exact result within all numerical prefactors.
In contrast, in the large-sample limit $N_{\Omega }\longrightarrow
\infty$, when interaction constant $v$ is finite and nonzero, one of
the terms becomes of the order of $1/N_{\Omega }$ compared to
another one, so that it can be dropped as an underextensive
contribution, while the remaining term again gives a correct result
within all numerical prefactors. Such a very subtle interplay
between the two terms indicates that the suggested formula for the
condensation energy must remain accurate not only in the considered
limits.

Actually, by manipulating the single-pair binding energy, which must
be determined numerically, we circumvent the problem of
inapplicability of large-$N$ approaches to the normal state, pointed
out by Richardson in Ref.~\cite{Rich3}. This difficulty can be
traced back to the non-analytic dependence of condensation energy on
$V$ in this limit, while in the limit of a very weak interaction it
is simply proportional to $V$. Moreover, in the first case, this
energy is an extensive quantity, while in the second one, it becomes
intensive. It is therefore challenging to unify both regimes within
a single self-consistent formalism.

In order to explore the applicability range of the obtained
expression of the condensation energy, we perform a systematic
numerical solution of the full set of Richardson equations, as given
by Eq. (\ref{Richardson}), for various values of $N$ from 1 to 50
and $v$ at the half-filling. Then, we compare the obtained numerical
results with the prediction of Eq. (\ref{final-cond}) (where $e_{1}$
is obtained from the numerical solution of Eq.~(\ref{Cooper})). We
also calculate the condensation energy by using the standard
grand-canonical BCS theory. The only difference with the common
version of this theory is the fact that we do not replace sums by
integrals when solving the gap equation and also when calculating
the condensation energy itself, which is of importance for systems
with relatively small number of pairs, since these replacements are
responsible for additional inaccuracies.

In the general case, the Richardson equations can be numerically
solved by Newton-Raphson method with a good initial guess. An exact
solution $R_j=2\varepsilon _{F_{0}}+2j/\rho\ (n=0,1,..,N-1)$ is a
solution for $v=0$. Therefore, we start with such initial values and
then find solutions with increasing $v$. In order to avoid the
singularity, new variables $\lambda_+,\lambda_-$ are
introduced\cite{Numer1}. When $v$ is close to the critical $v_c$,
the Newton-Raphson result does not converge to the solution if it
starts from the other side of the singularity\cite{Numer2}.
Therefore, an extrapolation step is taken for the new $v$ (close to
$v_c$), as proposed in Ref. \cite{Numer2}.

\begin{figure}[btp]
\begin{center}
\includegraphics*[width=7.5cm]{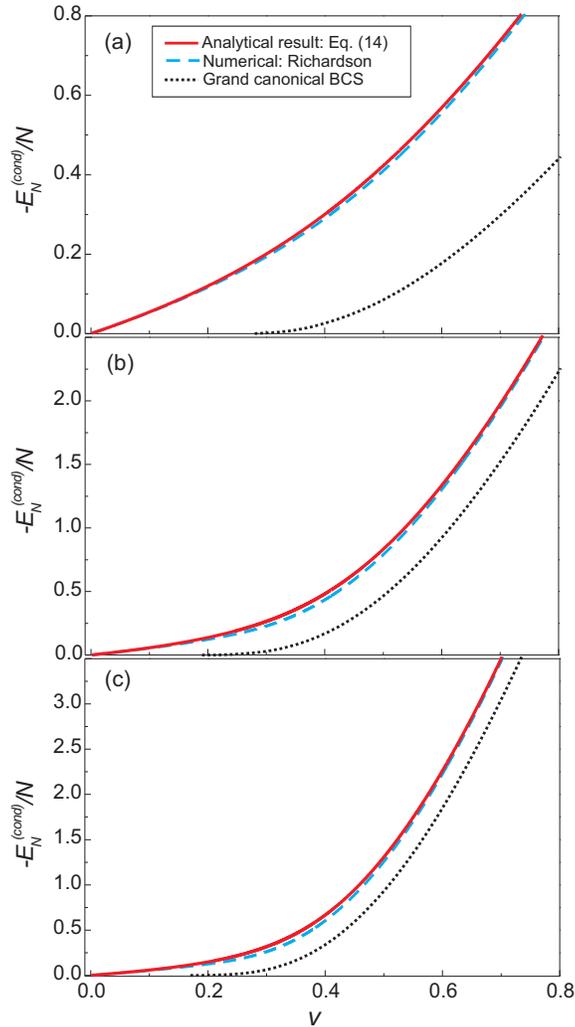}
\end{center}
\vspace{-0.5cm} \caption{ (Color online) The condensation energy per
pair as a function of the interaction constant, for $N$ pairs: $N=5$
(a), $N=25$ (b), and $N=50$ (c). Red (gray) solid lines show the
prediction according to our analytical formula (\ref{final-cond}),
blue (light gray) dashed lines represent the results of the
numerical solution of the full system of Richardson equations, black
dotted lines correspond to the grand-canonical BCS result. }
\end{figure}

The results are presented in Fig.~1 for three particular values of
$N$. The condensation energies are measured in terms of $2/\rho$.
Fig. 1(a -- c) give the dependence of the condensation energy per
pair, as a function of $v$, for $N=5$, 25, and 50, respectively.
Solid curves yield our prediction, dashed curves represent results
of the numerical solution of the Richardson equations, and dotted
lines are the grand-canonical BCS results. We see that there is
generally a good agreement between the numerical results and those
obtained from our formula. The similar agreement has been found for
other values of $N$. Therefore, our conjecture is justified. In
contrast, BCS results become accurate in the large-$N$ limit only:
as it is known, there is a range of small $v$ for any $N$, when the
grand-canonical BCS theory is \emph{qualitatively} incorrect, since
it predicts a disappearance of superconducting correlations (see
Fig.~1). In this fluctuation-dominated regime, our approach works
well. We found that the largest relative errors for the three cases
illustrated in Fig. 1 are $3$, $15$, and $23$ percent, respectively.
Thus, our approach is more efficient for small $N$, this case being
more interesting due to limitations of BCS theory for systems with
small number of pairs. We also would like to note that the accuracy
can, in principle, be improved by considering more terms in Eq.
(\ref{alfa-sums}) and using two-pairs "boundary condition" (and
possibly other configurations with even more pairs), although such a
procedure would lack simple analytical results, in contrast to the
present approach.

Within our approach, a single-pair binding energy $-e_{1}$ plays a
very important role, although we deal with the many-pair system. It
provides an energy scale, which is alternative to the
superconducting gap $\Delta$. Interestingly, the existence of an
additional scale for finite systems was revealed some time ago in
Ref. \cite{Schechter}, where it was shown that BCS results for the
ground state energy become inadequate already for level spacings
$1/\rho\approx2\Delta^{2}/\Omega$, which are much smaller than
$\Delta$ at $v\ll1$. Let us point out that, as easily seen by a
direct comparison, in the large-sample limit, $2\Delta^{2}/\Omega$
is nothing but $-e_{1}$ when $v$ is small. It is very unlikely that
such a coincidence is accidental. Therefore, we believe that the
additional energy scale found in Ref. \cite{Schechter} is connected
to the single-pair binding energy. Note that it was recently
argued~\cite{JETPLett} that standard BCS results for the
condensation energy in the thermodynamical limit can be also
interpreted in a simple way through the single-pair binding energy
and not through $\Delta$, as usual. In particular, in this limit,
$-e_{1}$ and $\Delta$ have similar, but different dependencies on
$v$, since $-e_{1}\sim\exp(-2/v)$, while $\Delta\sim\exp(-1/v)$ at
$v\ll1$, as discussed in detail in Ref. \cite{JETPLett}.

It is perspective to extend our analysis to excited states. We also
think that, probably, similar ideas can be applied to other types of
energy-level distributions, in particular, to those, which are
characterized by the inversion symmetry around the middle point of
the band. Furthermore, the analysis of the electron-hole symmetries
can be helpful in treatments of other Bethe-ansatz equations, among
which Richardson equations are just one of the examples.

Notice that the derived \emph{exact} relation for the condensation
energy, given by Eq.~(\ref{switch-cond}), can be used as a test to
check the accuracy of the numerically-found solutions of Richardson
equations.

\section{Conclusions}

We suggested a simple expression for the ground state energy of the
pairing Hamiltonian for the case of the equally-spaced model along
the crossover from the superconducting regime to the pairing
fluctuation regime. This expression is derived from the peculiar
electron-hole symmetry of the pairing Hamiltonian and relies on the
conjecture, which enables us to reduce the task to the one-pair
problem. The electron-hole symmetry is encoded in the Hamiltonian,
but it is hidden in Richardson equations, which provide an exact
many-body solution of the problem. The obtained expression of the
ground state energy depends on the binding energy of a single pair,
which, in the general case, must be determined numerically by
solving a single Richardson equation. The latter problem is much
simpler than the solution of the full set of Richardson equations.
This quantity also provides an additional energy scale associated
with superconducting correlations, which seems to be connected with
the energy scale revealed in Ref. \cite{Schechter}.

The comparison with the results of the full numerical resolution of
Richardson equations demonstrated a generally good accuracy of the
suggested formula, while the usual grand-canonical BCS approach
fails even qualitatively in the fluctuation-dominated regime. The
accuracy is better in the case of the system with small number of
pairs, which is of particular interest due to limitations of BCS
method in this situation.


\begin{acknowledgments}

This work was supported by the ``Odysseus'' Program of the Flemish
Government and the Flemish Science Foundation (FWO-Vl).
W.V.P. acknowledges useful discussions with Monique Combescot
and
the support from the Dynasty Foundation,
the RFBR (project no. 12-02-00339), and RFBR-CNRS programme
(project no. 12-02-91055).
\end{acknowledgments}


\begin{references}

\bibitem{Rich1}
    R. W. Richardson, Phys. Lett. \textbf{3} (1963) 277.

\bibitem{Rich3}
    R. W. Richardson, J. Math. Phys. \textbf{18} (1977) 1802;
    M. Gaudin, J. Phys. (Paris) \textbf{37} (1976) 1087.

\bibitem{Braun}
    K. Dietrich, H. J. Mang, and J. H. Pradal,
    Phys. Rev. B \textbf{22} (1964) 135;
    F. Braun and J. von Delft,
    Phys. Rev. Lett. \textbf{81} (1998) 4712.

\bibitem{Bogoliubov1}
    N. N. Bogoliubov, Usp. Fiz. Nauk \textbf{67} (1959) (549)
    [Sov. Phys. Usp. \textbf{2} (1959) 236].

\bibitem{Pogosyan}
    J. von Delft and R. Poghossian,
    Phys. Rev. B \textbf{66} (2002) 134502.

\bibitem{Dukel}
    J. Dukelsky, S. Pittel, and G. Sierra,
    Rev. Mod. Phys. \textbf{76} (2004) 643.

\bibitem{Dukel1}
    N. Sandulescu, B. Errea, and J. Dukelsky,
    Phys. Rev. C \textbf{80} (2009) 044335.

\bibitem{Belzig}
    S. Staudenmayer, W. Belzig, and C. Bruder,
    Phys. Rev. A \textbf{77} (2008) 013612.

\bibitem{Gladilin1}
    J. Tempere, V. N. Gladilin, I. F. Silvera, and J. T. Devreese, Phys. Rev. B \textbf{72} (2005) 094506.

\bibitem{Osterloh}
    L. Amico and A. Osterloh,
    Phys. Rev. Lett. \textbf{88} (2002) 127003;
    H.-Q. Zhou, J. Links, R. H. McKenzie, and M. D. Gould,
    Phys. Rev. B \textbf{56} (2002) 060502(R);
    G. Gorohovsky and E. Bettelheim,
    Phys. Rev. B \textbf{84} (2011) 224503.

\bibitem{Kurland}
    I. L. Kurland, I. L. Aleiner, and B. L. Altshuler,
    Phys. Rev. B \textbf{62} (2000) 14886.

\bibitem{Aleiner}
    I. L. Aleiner, P. W. Brouwer, and L. I. Glazman,
    Phys. Rep. \textbf{358} (2002) 309.

\bibitem{Ralph}
    J. von Delft and D. C. Ralph,
    Phys. Rep. \textbf{345} (2001) 61.

\bibitem{Vinokur}
    I. S. Beloborodov, A. V. Lopatin, V. M. Vinokur, and K. B. Efetov, Rev. Mod. Phys. \textbf{79} (2007) 469.

\bibitem{Geyer}
    I. Snyman and H. B. Geyer,
    Phys. Rev. B \textbf{73} (2006) 144516.

\bibitem{Eagles}
    D. M. Eagles, Phys. Rev. \textbf{186} (1969) 456.

\bibitem{Ambeg}
    R. A. Smith and V. Ambegaokar,
    Phys. Rev. Lett. \textbf{77} (1996) 4962.

\bibitem{Dussel}
    G. Sierra, J. Dukelsky, G. G. Dussel, J. von Delft, and F. Braun, Phys. Rev. B \textbf{61} (2000) R11890.

\bibitem{Gorkov}
    L. P. Gor'kov and G. M. Eliashberg,
    Zh. Eksp. Theor. Fiz. \textbf{48} (1965) 1407
    [Sov. Phys. JETP \textbf{21} (1965) 940];
    K. B. Efetov,
    Adv. Phys. \textbf{32} (1983) 53.

\bibitem{We}
    M. Crouzeix and M. Combescot,
    Phys. Rev. Lett. \textbf{107} (2011) 267001;
    W. V. Pogosov,
    J. Phys.: Condens. Matter \textbf{24} (2012) 075701.

\bibitem{Schechter}
    M. Schechter, Y. Imry, Y. Levinson, and J. von Delft,
    Phys. Rev. B \textbf{63} (2001) 214518.

\bibitem{Altshuler}
    E. A. Yuzbashyan, A. A. Baytin, and B. L. Altshuler,
    Phys. Rev. B \textbf{71} (2005) 094505.

\bibitem{Numer1}
    A. Faribault, P. Calabrese, and J. S. Caux,
    J. Math. Phys. \textbf{50} (2009) 095212.

\bibitem{Numer2}
    S. Rombouts, D. Van Neck, and J. Dukelsky,
    Phys. Rev. C \textbf{69} (2004) 061303(R).

\bibitem{JETPLett}
    W. V. Pogosov and M. Combescot,
    Pis'ma v ZhETF \textbf{92} (2010) 534
    [JETP Lett. \textbf{92} (2010) 534].



\end{references}
\end{document}